\documentclass{article}
\usepackage{spconf,amsmath,graphicx}
\usepackage{caption,subcaption,pgfplots}
\usepackage{hyperref}

\title{BIRD: Big Impulse Response Dataset}
\name{Fran\c{c}ois Grondin\textsuperscript{1}, Jean-Samuel Lauzon\textsuperscript{1}, Simon Michaud\textsuperscript{1}, Mirco Ravanelli\textsuperscript{2}, Fran\c{c}ois Michaud\textsuperscript{1}}
\address{\textsuperscript{1}IntRoLab - Universit\'e de Sherbrooke, \textsuperscript{2}Mila - Universit\'e de Montr\'eal\\ \small\texttt{\{francois.grondin2,jean-samuel.lauzon,simon.michaud,francois.michaud\}@usherbrooke.ca} \\ \small\texttt{mirco.ravanelli@gmail.com}}
\begin{document}
%\ninept
%
\maketitle
\begin{abstract}
This paper introduces BIRD, the Big Impulse Response Dataset.
This open dataset consists of 100,000 multichannel room impulse responses (RIRs) generated from simulations using the Image Method, making it the largest multichannel open dataset currently available.
These RIRs can be used to perform efficient online data augmentation for scenarios that involve two microphones and multiple sound sources. The paper also introduces use cases to illustrate how BIRD can perform data augmentation with  existing speech corpora.
\end{abstract}
\begin{keywords}
Room Impulse Response, Data augmentation, Reverberation, Speech
\end{keywords}
\section{Introduction}
\label{sec:intro}

Distant speech recognition remains a challenging task as the target speech usually gets corrupted by reverberation and background noise \cite{tang2018study}.
There is often a difference between real-life scenarios and the clean recording conditions, as most speech datasets are recorded  with a single close-talking microphone (e.g., LibriSpeech \cite{panayotov2015librispeech}, TIMIT \cite{zue1990speech}, WSJ \cite{paul1992design}, etc.)
To reduce domain mismatch, it is common to capture audio using multiple microphones at test time, and use methods such as Minimum Variance Distortionless Response (MVDR) \cite{habets2009new, erdogan2016improved} and Generalized Eigenvalue Decomposition (GEV) \cite{heymann2015blstm} beamforming to enhance speech prior to performing recognition.
These multichannel enhancement methods often rely on time-frequency masks estimated using neural networks.
Datasets composed of multichannel recordings in reverberant conditions therefore become handy to train these neural networks.

The Aachen Impulse Response (AIR) dataset \cite{jeub2009binaural} provides recording using two microphones in a real environment, with a fixed spacing between both microphones and six room configurations. Similarly, the ACE Corpus \cite{eaton2015ace} provides recording with a phone, a notebook and a 32-channel spherical microphone array in seven different rooms.
The DIRHA project dataset \cite{ravanelli2015dirha}, VoiceHome corpus \cite{bertin2016french} and Sweet-Home \cite{vacher2014sweet} corpus also provide recordings with microphone arrays in different rooms.
These recorded datasets are however limited to few microphone arrays and room configurations.

One alternative consists of recording or simulating the room impulse responses (RIRs), and then convolving them with the speech corpus of interest \cite{ravanelli15, stewart2010database}.
While recording real RIRs can be cumbersome \cite{ravanelli14}, simulating synthetic RIRs provides an alternative to augment speech with a large range of room configurations and microphone array shapes.
The Image Method \cite{allen1979image} offers a simple yet effective way to simulate reflections for a rectangular empty room.
Simulations can be done using a central processing unit (CPU) in the Matlab environment \cite{habets2006room} or using Python libraries such as PyRoomAcoustics \cite{scheibler2018pyroomacoustics}.
The computations can also be sped up with a graphics processing unit (GPU) \cite{fu2016gpu}.
However, simulating different RIRs for each sample while training a neural network uses precious computing resources, and can even act as a bottleneck.
Ko et al. \cite{ko2017study} present a dataset that includes both real and simulated RIRs, but these are limited to a single microphone.
In some cases, RIRs are generated and convolved using a specific speech corpus offline, and the augmented signals are then used for training \cite{hershey2016deep}.
This can however require a significant amount of storage, as opposed to performing data augmentation online while training.

This paper presents a new dataset named BIRD, for the Big Impulse Response Dataset.
This dataset is the largest open multichannel RIR corpus currently available.
It consists of 100,000 precomputed multichannel RIRs obtained using the Image Method, divided into 10 balanced folds of 10,000 RIRs each, that can be combined to generate training, validation, and testing sets.
Simulations are performed for a wide range of room dimensions and spacing between the microphones.
These RIRs can be loaded at training time and convolved on-the-fly with any audio corpus to generate augmented audio signals.
The paper also presents four scenarios in which the BIRD dataset is exploited for online data augmentation.

\section{Dataset}
\label{sec:format}

All RIRs in BIRD are generated using the implementation of the Image Method made publicly available online by Habets \cite{habets2006room}\footnote{\url{https://github.com/ehabets/RIR-Generator}}.
\begin{figure}[!ht]
    \centering
    \includegraphics[width=0.95\linewidth]{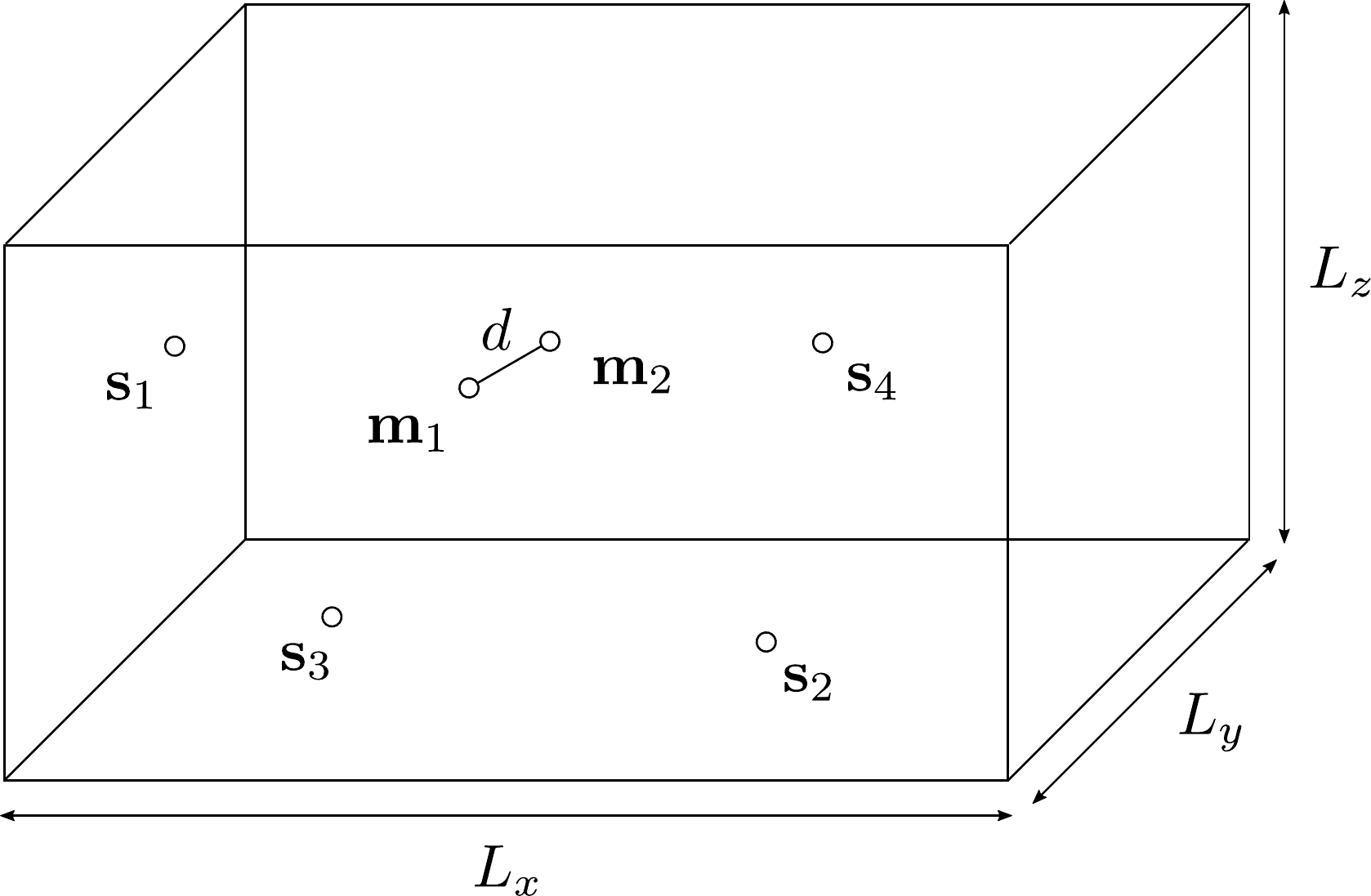}
    \caption{Room of dimensions $L_x \times L_y \times L_z$ meters with 4 sound sources and 2 microphones spaced by $d$ meters.}
    \label{fig:dataset_room}
\end{figure}

Figure \ref{fig:dataset_room} illustrates the setup used for the simulations, where two microphones and four sources are positioned randomly in the virtual room.
Each simulation is performed using a rectangular room of dimensions $L_x \times L_y \times L_z$ m.
These dimensions are chosen by uniformly sampling ranges denoted by $[L_{x,min},L_{x,max}]$, $[L_{y,min},L_{y,max}]$ and $[L_{z,min},L_{z,max}]$, which reflect realistic rooms in houses and offices:
\begin{equation}
    \mathbf{L} =
    \left[
    \begin{array}{c}
        L_x \\
        L_y \\
        L_z \\
    \end{array}
    \right] \sim 
    \left[
    \begin{array}{c}
        \mathcal{U}(L_{x,min},L_{x,max}) \\
        \mathcal{U}(L_{y,min},L_{y,max}) \\
        \mathcal{U}(L_{z,min},L_{z,max}) \\
    \end{array}
    \right],
    \label{eq:L}
\end{equation}
where $\mathcal{U}(a,b)$ stands for a uniform distribution within the interval $[a,b]$.

The absorption of the acoustic energy on the walls, ceiling, and floor depends on the type of surfaces, and is usually modeled by an absorption coefficient.
A coefficient of one represents perfect acoustic absorption (anechoic), whereas a value of zero stands for perfect reflection without attenuation.
In this dataset, we assume the absorption coefficient $\alpha$ is identical for all surfaces, and lies within the uniform range $[\alpha_{min}, \alpha_{max}]$, with $\alpha \sim \mathcal{U}(\alpha_{min}, \alpha_{max})$.
The speed of sound $c$ (in m/sec) varies mainly with temperature.
It therefore lies in a uniform range that can match the typical indoor temperatures, such that $c \sim \mathcal{U}(c_{min}, c_{max})$.

Sound sources are then positioned randomly in the room.
They are assumed to be omnidirectional, and the positioning ensures there is a minimum distance of $0.5$ m between each source and the surrounding surfaces as follows:

\begin{equation}
    \mathbf{s}_i = \left[
    \begin{array}{c}
        s_{i,x} \\
        s_{i,y} \\
        s_{i,z} \\
    \end{array}
    \right] \sim
    \left[
    \begin{array}{c}
        \mathcal{U}(0.5, L_x - 0.5) \\
        \mathcal{U}(0.5, L_y - 0.5) \\
        \mathcal{U}(0.5, L_z - 0.5) \\
    \end{array}
    \right].
\end{equation}

Similarly, the pair of microphones is positioned randomly in the room and centered at:
\begin{equation}
    \mathbf{o} = \left[
    \begin{array}{c}
    o_x \\
    o_y \\
    o_z \\
    \end{array}
    \right] \sim 
    \left[
    \begin{array}{c}
    \mathcal{U}(0.5, L_x - 0.5) \\
    \mathcal{U}(0.5, L_y - 0.5) \\
    \mathcal{U}(0.5, L_z - 0.5) \\
    \end{array}
    \right],
\end{equation}
where the distance $d$ between both microphones is uniformly sampled with $d \sim \mathcal{U}(d_{min}, d_{max})$.

To cover a broad range of orientations, the pair of microphones rotates with random yaw ($\theta_{yaw} \sim \mathcal{U}(0, 2\pi)$), pitch ($\theta_{pitch} \sim \mathcal{U}(0, 2\pi)$) and roll ($\theta_{roll} \sim \mathcal{U}(0, 2\pi)$), where the rotation matrices $\mathbf{R}_x(\theta_{yaw})$, $\mathbf{R}_y(\theta_{pitch})$ and $\mathbf{R}_z(\theta_{roll})$ correspond to:
\begin{equation}
    \mathbf{R}_x(\theta_{roll}) = 
    \left[ 
    \begin{array}{ccc}
    1 & 0 & 0 \\
    0 & \cos(\theta_{roll}) & -\sin(\theta_{roll}) \\
    0 & \sin(\theta_{roll}) & \cos(\theta_{roll}) \\
    \end{array}
    \right],
\end{equation}
\begin{equation}
    \mathbf{R}_y(\theta_{pitch}) = 
    \left[ 
    \begin{array}{ccc}
    \cos(\theta_{pitch}) & 0 & \sin(\theta_{pitch}) \\
    0 & 1 & 0 \\
    -\sin(\theta_{pitch}) & 0 & \cos(\theta_{pitch}) \\
    \end{array}
    \right],
\end{equation}
\begin{equation}
    \mathbf{R}_z(\theta_{yaw}) = 
    \left[ 
    \begin{array}{ccc}
    \cos(\theta_{yaw}) & -\sin(\theta_{yaw}) & 0 \\
    \sin(\theta_{yaw}) & \cos(\theta_{yaw}) & 0 \\
    0 & 0 & 1 \\
    \end{array}
    \right].
\end{equation}

The microphones are positioned in the room according to:
\begin{equation}
    \mathbf{m} = \mathbf{R}_x(\theta_{roll}) \mathbf{R}_y(\theta_{pitch}) \mathbf{R}_z(\theta_{yaw})\mathbf{u} +
    \mathbf{o},
\end{equation}
where $\mathbf{u}$ stands for the pair of microphone positioned parallel to the $x$-axis:
\begin{equation}
    \mathbf{u} = 
    \left[
    \begin{array}{cc}
    \mathbf{u}_1 & \mathbf{u}_2
    \end{array}
    \right] = 
    \left[
    \begin{array}{cc}
    -d/2 & +d/2 \\
    0 & 0 \\
    0 & 0 \\
    \end{array}
    \right],
\end{equation}
and $\mathbf{m}$ represents the absolute position of both microphones in the room:
\begin{equation}
    \mathbf{m} = \left[ 
    \begin{array}{cc}
    \mathbf{m}_1 & \mathbf{m}_2
    \end{array}
    \right] = 
    \left[
    \begin{array}{cc}
    m_{1,x} & m_{2,x} \\
    m_{1,y} & m_{2,y} \\
    m_{1,z} & m_{2,z} \\
    \end{array}
    \right].
\end{equation}

Table \ref{tab:parameters} presents the intervals used for generating randomly the simulation parameters.

\begin{table}[!ht]
    \centering
    \caption{Simulation parameters}
    \vspace{-5pt}
    \renewcommand{\arraystretch}{1.2}
    \begin{tabular}{|c|c|c|c|c|}
        \cline{1-2} \cline{4-5}
        Parameter & Value & & Parameter & Value \\
        \cline{1-2} \cline{4-5}
        $L_{x,min}$ & $5.0$ & & $\alpha_{min}$ & $0.2$ \\
        $L_{x,max}$ & $15.0$ & & $\alpha_{max}$ & $0.8$ \\
        $L_{y,min}$ & $5.0$ & & $c_{min}$ & $340.0$ \\
        $L_{y,max}$ & $15.0$ & & $c_{max}$ & $355.0$ \\
        $L_{z,min}$ & $3.0$ & & $d_{min}$ & $0.01$ \\
        $L_{z,max}$ & $4.0$ & & $d_{max}$ & $0.30$ \\
        \cline{1-2} \cline{4-5}
    \end{tabular}
    \label{tab:parameters}
\end{table}

The Image Method generates an impulse response for each microphone and source, denoted as $\hat{h}_{i,k}[n]$, where $k \in \{1,2\}$ stands for the microphone index, $i \in \{1,2,3,4\}$ for the source index and $n \in \{1,2,\dots,N\}$ for the sample index.
The maximum absolute value for all microphones, sources and samples are calculated as follows:
\begin{equation}
    A_{max} = \max_{k \in \{1,2\}, i \in \{1,2,3,4\}, n \in \{1,2,\dots,N\}}{|h_{i,k}[n]|},
\end{equation}
and then scale the amplitude such that the RIRs span $99\%$ of the amplitude range:
\begin{equation}
    h_{i,k}[n] = \left(\frac{0.99}{A_{max}}\right) \hat{h}_{i,k}[n].
\end{equation}

\begin{figure}[!ht]
    \centering
    \begin{subfigure}[b]{\linewidth}
        \centering
        \includegraphics[width=\linewidth]{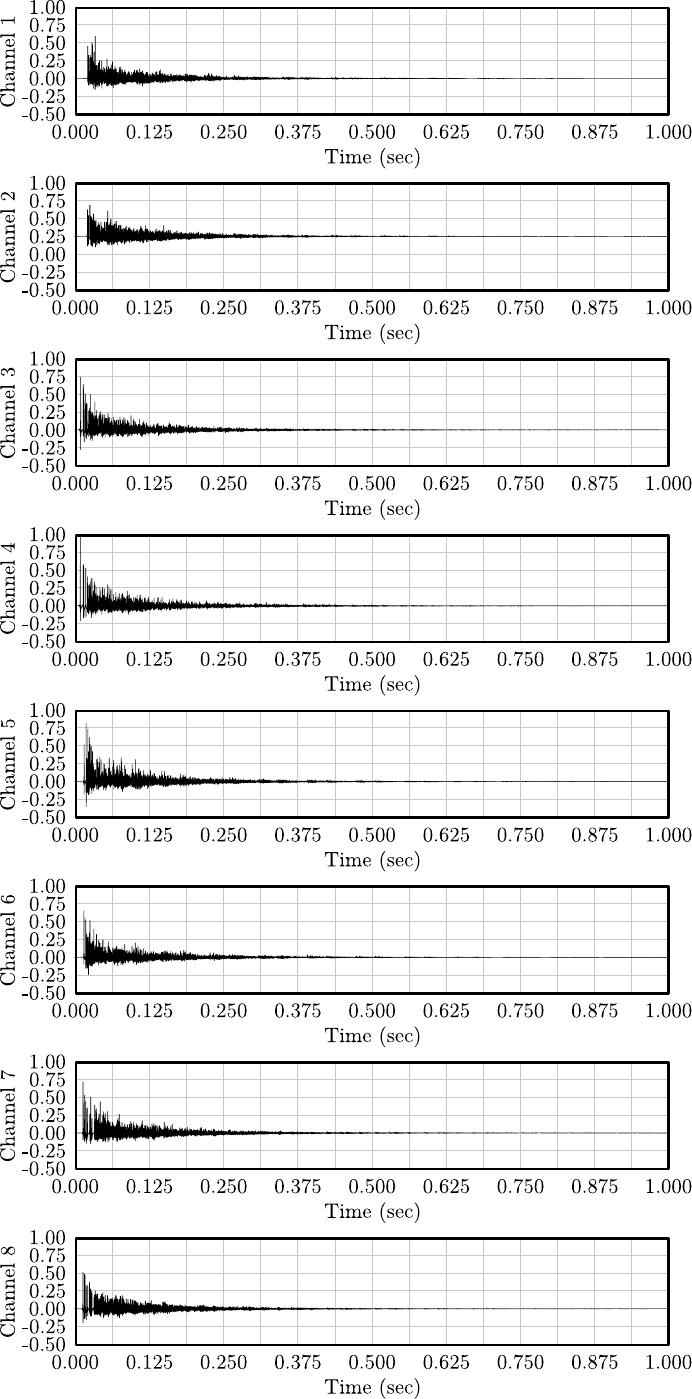}
    \end{subfigure}
    \begin{subfigure}[b]{\linewidth}
    \vspace{3pt}
    \centering\fbox{\parbox{0.95\linewidth}{\texttt{\scriptsize\{
    "L": [14.83, 11.49, 3.01], 
    "alpha": 0.36,
    "c": 350.5,
    "mics": [[14.141, 2.934, 1.895], 
             [14.224, 3.010, 2.161]],
    "srcs": [[0.811, 5.702, 1.547], 
             [6.658, 4.000, 2.582], 
             [5.340, 9.433, 1.775], 
             [12.164, 8.109, 2.161]]
    \}
    }}}
    \end{subfigure}
    \vspace{-5pt}
    \caption{Example of a FLAC file in BIRD. The RIRs are stored in the audio channels 1-8, and the simulation parameters are encoded as a JSON string and stored in the metadata section of the file.}
    \label{fig:flac}
\end{figure}

This scaling operation ensures there is no clipping and that the distortion introduces with sample discretization is minimal.
Each impulse response is saved in a Free Lossless Audio Codec (FLAC) file\footnote{\url{https://xiph.org/flac/}} with 16-bit precision, where channel $1$ corresponds to $h_{1,1}$, channel $2$ to $h_{1,2}$, channel $3$ to $h_{2,1}$, and so on until channel $8$ to $h_{4,2}$.
Simulations are performed with a sample rate of 16,000 samples/sec, and each RIR lasts 1 sec (16,000 samples).
The metadata of each file also contains the RIR simulation parameters in JSON format.
Figure \ref{fig:flac} shows an example of a FLAC file structure.

The dataset is split into a total of 10 balanced folds, each containing 10,000 RIRs.
The dataset and some examples based on the Pytorch framework are available online\footnote{\url{https://pybird.io}}.

\section{Data augmentation}

A wide range of audio mixtures can be generated using the BIRD dataset.
A mixture of sound sources is defined as follows:
\begin{equation}
    y_k = v\sum_{i=1}^{I}{g_i y_{i,k}},
\end{equation}
where each source-microphone combination is computed using linear convolution, i.e. $y_{i,k} = h_{i,k}\,\ast\,x_{i}$.
The variable $x_i$ stands for the clean audio signal that corresponds to source $i$ and comes from any speech or non-speech corpus (e.g., LibriSpeech \cite{panayotov2015librispeech}, TIMIT \cite{zue1990speech}, WSJ \cite{paul1992design}, ESC50 \cite{piczak2015esc}, etc.)
The gains $g_i$ are introduced to change the signal-to-interference-plus-noise ratio (SINR) for each source, and $v$ controls the volume level of the mixture.

The metadata also provide useful information.
For instance, the time difference of arrival (TDOA) (in samples) for source $i$ can be retrieved as follows:
\begin{equation}
    \tau_{i} = \frac{f_S}{c}(\mathbf{m}_1 - \mathbf{m}_2) \cdot \left(\frac{\mathbf{s}_i - \frac{1}{2}(\mathbf{m}_1 + \mathbf{m}_2)}{\lVert \mathbf{s}_i - \frac{1}{2}(\mathbf{m}_1 + \mathbf{m}_2)\rVert_2}\right),
    \label{eq:tdoa}
\end{equation}
where $f_S$ stands for the sample rate (samples/sec), $\lVert\dots\rVert_2$ the Euclidean distance, and $(\cdot)$ the dot product.
It is also possible to estimate the reverberation time ($RT_{60}$) (in sec) using the Sabin-Franklin equation \cite{pierce2019acoustics}:
\begin{equation}
    RT_{60} = \left(\frac{12 \ln 10}{\alpha c}\right)\left(\frac{L_x L_y L_z}{L_x L_y + L_y L_z + L_z L_x}\right)
    \label{eq:rt60}
\end{equation}

Ideal ratio masks (IRMs) are often estimated to separate the source $i$, and can be computed as follows \cite{narayanan2013ideal}:
\begin{equation}
    M_{i,k}[f] = \frac{\lVert Y_{i,k}[f] \rVert^2_2}{\sum_{i=1}^{I}{\lVert Y_{i,k}[f] \rVert^2_2}},
    \label{eq:M}
\end{equation}
where $Y_{i,k}[f]$ and $Y_k[f]$ stand for the Short-Time Fourier Transform (STFT) of $y_{i,k}$ and $y_k$, respectively, and $f$ stands for the frequency bin index.

Figures \ref{fig:tdoas} and \ref{fig:rt60s} show the distributions of the TDOAs and RT60s obtained from the BIRD dataset.
These histograms suggest that the TDOAs and RT60s follow a Laplace and a gamma distributions, respectively.
\begin{figure}[!ht]
    \centering
    \begin{tikzpicture}
    \begin{axis}[ybar interval, ymajorticks=false, xticklabel style={rotate=90, font=\scriptsize, /pgf/number format/.cd, fixed, print sign}, width=1.1\linewidth, height=0.5\linewidth, grid=none, xtick style={draw=none}]
    \addplot table [x=tdoa, y=count, col sep=comma] {tdoas.csv};
    \end{axis}
    \end{tikzpicture}
    \vspace{-5pt}
    \caption{Distribution of TDOAs (in samples)}
    \label{fig:tdoas}
\end{figure}
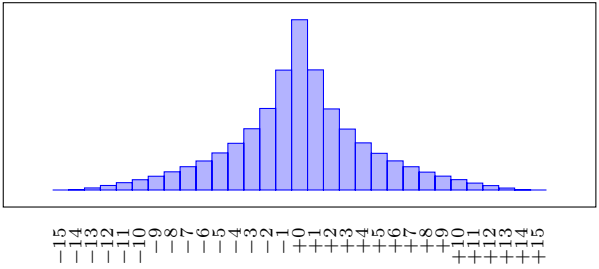
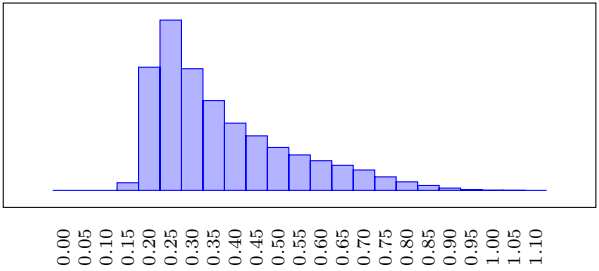
\begin{figure}[!ht]
    \centering
    \begin{tikzpicture}
    \begin{axis}[ybar interval, ymajorticks=false, xticklabel style={rotate=90, font=\scriptsize, /pgf/number format/.cd, fixed, zerofill, precision=2}, width=1.1\linewidth, height=0.5\linewidth, grid=none, xtick style={draw=none}]
    \addplot table [x=rt60, y=count, col sep=comma] {rt60s.csv};
    \end{axis}
    \end{tikzpicture}
    \vspace{-5pt}
    \caption{Distribution of RT60s (in sec)}
    \label{fig:rt60s}
\end{figure}

Numerous scenarios can be simulated and used when training a neural network to perform prediction.
Here are a few examples:

\begin{itemize}
    
    \item \textbf{Sound source localization}: When given as an input the spectrogram of the mixture obtained for two active sound sources, the neural network estimates the TDOAs obtained with (\ref{eq:tdoa}) that corresponds to both sources (similar to \cite{grondin2019sound}):
    \begin{equation}
        \{Y_1, Y_2\} \rightarrow \{\hat{\tau}_1, \hat{\tau_2}\}
    \end{equation}
    
    \item \textbf{Reverberation time estimation}: When given as an input the spectrogram of the mixture, the model estimates the reverberation time ($RT_{60}$) obtained in (\ref{eq:rt60}) (as in \cite{lee2019ensemble}):
    \begin{equation}
        \{Y_1, Y_2\} \rightarrow \hat{RT}_{60}
    \end{equation}
    
    \item \textbf{Counting speech sources}: When given as an input the spectrogram of the mixture, the network estimates the number of active sources (similar to what is done in \cite{stoter2018countnet}):
    \begin{equation}
        \{Y_1, Y_2\} \rightarrow \hat{I}
    \end{equation}
    
    \item \textbf{Ideal Ratio Mask estimation}: When given as an input the spectrogram of the mixture and a target TDOA, the model estimates the IRM obtained in (\ref{eq:M}) (as in \cite{grondin2020gev}):
    \begin{equation}
        \{Y_1, Y_2, \tau\} \rightarrow \{\hat{M}_1, \hat{M}_2\}
    \end{equation}
    
\end{itemize}

We provide code in the BIRD repository that implements augmented datasets combining BIRD and LibriSpeech, which are compatible with the PyTorch data loader module.
The BIRD dataset itself follows the architecture of other datasets provided with the Torch audio package\footnote{\url{https://pytorch.org/audio/}}, and all data can be directly downloaded and saved to disk while instantiating the class.

\section{Conclusion}

This paper presents the BIRD dataset, the largest multichannel RIR corpus currently available, and illustrates how this dataset can be used with a clean speech corpus to perform data augmentation for multi-microphone scenarios with multiple sound sources.
The code is provided online to easily integrate BIRD to machine learning projects using the PyTorch framework.

In future work, we would like to extend the dataset to incorporate simulated RIRs that match the geometries of the most popular commercially available microphone arrays.
We could also simulate rooms with complex geometries (i.e., go beyond rectangular rooms), and select different absorption coefficients for each surface that models different types of materials (e.g., carpet, dry walls, concrete, suspended ceiling, etc.)

\bibliographystyle{IEEEbib}
\bibliography{refs}

\end{document}